\documentclass[12pt,a4paper]{article}
\usepackage[utf8]{inputenc}
\usepackage[french,english]{babel}
\usepackage{authblk}
\usepackage{multicol}
\usepackage{hyperref}
\hypersetup{
  colorlinks = true,
  linkcolor = red,
  citecolor  = red,
  urlcolor = blue
}

\usepackage{graphicx}
\usepackage{listings}
\usepackage{amsmath}
\usepackage{array}

\newcommand{\fig}[1]{figure~\ref{#1}}

\newcommand{\keywords}[1]{\textbf{Keywords:}~#1}

\title{Characterization of Android malware based on opcode analysis}
\author[1]{Alain Menelet}
\author[2]{Charles-Edmond Bichot}
\affil[1]{Armée de l'air, ESIOC, France}
\affil[2]{Armée de l’air, CNRS, LIRIS, École de l'air, CREA, France}

\date{June 2020}

\begin{document}

\maketitle

\begin{abstract}
The Android operating system is the most spread mobile platform in the world. Therefor attackers are producing an incredible number of malware applications for Android. Our aim is to detect Android's malware in order to protect the user. To do so really good results are obtained by dynamic analysis of software, but it requires complex environments. In order to achieve the same level of precision we analyze the machine code and investigate the frequencies of ngrams of opcodes in order to detect singular code blocks. This allow us to construct a database of infected code blocks. Then, because attacker may modify and organized differently the infected injected code in their new malware, we perform not only a semantic comparison of the tested software with the database of infected code blocks but also a structured comparison. To do such comparison we compute subgraph isomorphism. It allows us to characterize precisely if the tested software is a malware and if so in witch family it belongs. Our method is tested both on a laboratory database and a set of real data. It achieves an almost perfect detection rate.
\end{abstract}



\keywords{malware; malware detection; malware characterization; malware families; android; static analysis; opcodes, subgraph isomorphism}

\maketitle

\section{State of the art}

Android has captured roughly 78,8\% of today’s smartphone market with Version 6.0 of its operating system installed on over 28,1\% of existing equipment. Over 99\% of malware designed to target mobile devices are aimed at Android devices, and they keep increasing in number and ingenuity~\cite{fsecuresecurity2017}. 
They resort to such mechanisms as obfuscation, encryption, or communication with their command and control center in order to evade detection systems. 

In early 2017, Google launched Verify Apps, a protection system that monitors and profiles the behavior of Android applications. If an application is flagged as potentially harmful, Verify Apps immediately informs the user that it needs to be removed. According to data by Google, Verify Apps could analyze more than 750 million applications per day and, although no detailed information has been released, it is likely to perform both static and dynamic security verification. Experts have shown that applications verified in Google Play were run in a virtual machine, whose profile can be reconstructed~\cite{Oberheide}. However, this protection system is far from infallible, as evidenced by malware Dvmap~\cite{dvmap}, which after gaining root access was able to deactivate Verify Apps and go undetected.

Malware detection is generally done through the use of signatures, based on apks or behavioral profiles. Signatures and profiles can be built based on a static or dynamic analysis of the malware. Static analysis allows to build malware signatures against which the signature of a new sample can be compared~\cite{griffin2009automatic,feng2014apposcopy}. 

The method we put forward aims at describing the behavioral profile of an Android application. Behavioral analysis is often done using a dynamic approach. However, we consider that static analysis provides an appropriate representation of this profile, by taking into account all possible junctions, contrary to dynamic analysis that could be bypassed by malware~\cite{Abraham2015}. Indeed, malware can carry out evasive techniques like transformation in order to evade signatures~\cite{Rastogi2014} or sandbox detection~\cite{Vidas:2014:EAR}. As a result, we offer a static approach based on the bytecode blocks constitutive of the application, with a control flow graph built from possible sequences between these blocks.

Other Android malware detection methods based on the API exist. In~\cite{MaMaDroid}, Mariconti \emph{and all} have studied the API call sequences in order to determine the behavior of the application. To this end, they proceeded with the extraction of call graphs and used the Markov chain to represent the call sequences. They came up with an interesting detection rate of 99\% for the new malware from the same period as the malware used for the learning process. This value decreases as one moves away from the learning period. In~\cite{li2015detection}, Li \emph{and all} proceeded with the extraction of APIs in order to determine characteristic dependencies. Relying on automatic learning algorithms, especially using decision tree forests, they have managed to identify close to 96\% of the malware present in their database. This research shows the correlation between APIs and malware behavior -a correlation inherent in the Android architecture. In~\cite{gascon2013structural}, Gascon \emph{and all} presented a study on the relevance of function call graphs in the search for variants. Using automatic learning via support vector machines on over 12,000 pieces of malware, they achieved a detection rate of 89\% with only 1\% of false positives.

\section{Construction of the approach}

\subsection{Benefits of the proposed method}

This paper presents a novel approach to malware detection based on subgraph isomorphism using blocks of opcode. 
These opcodes correspond to the opcode present in the bytecode of the Dalvik virtual machine.
Resorting to the subgraph isomorphism problem allows to conduct a structured comparison between the opcode blocks on top of a semantic comparison. This is the first benefit of the present approach compared to the state of the art that only uses semantic comparison as a general rule.

Subgraph isomorphism allows to check unknown software against known malware for similarities. In other words, the subgraph isomorphism computing algorithm is going to compare the graphs of opcode blocks of the software that needs to be scanned to a database of graphs of malware opcode blocks. This is the second benefit of the present approach compared to previous research that often does not go into such level of detail, that is to say machine language.

Machine language is our focus and more specifically operation codes or “opcodes”. We assume that in any language the choice of words reveals the intentions of their user. In this regard, we consider that opcode blocks resulting from infected code feature a vocabulary that is slightly different from other opcode blocks. Our approach uses the relative frequency of opcode blocks following a text mining method. This is the third benefit of the present method, which analyzes machine language to extract similarity knowledge with infected code.

We make the basic assumption that malware developers follow the good practices of software engineering by reusing existing code to develop new malware. Identical blocks of machine code can be identified through the signature principle. This assumption especially holds for malware that combine clean programs with bits of infected code. For detection accuracy, one needs to keep in mind that malware developers may adapt, modify, complete, that is to say transform the bits of infected code they use. We put forward an approach which counters this obstacle. This is the fourth benefit of the present method, which is impervious can hold against withstand transformations to the infected code and is thereby able to detect pieces of malware from the same family.

\subsection{Overview of the present method}

\begin{figure}
  \includegraphics[width=\textwidth]{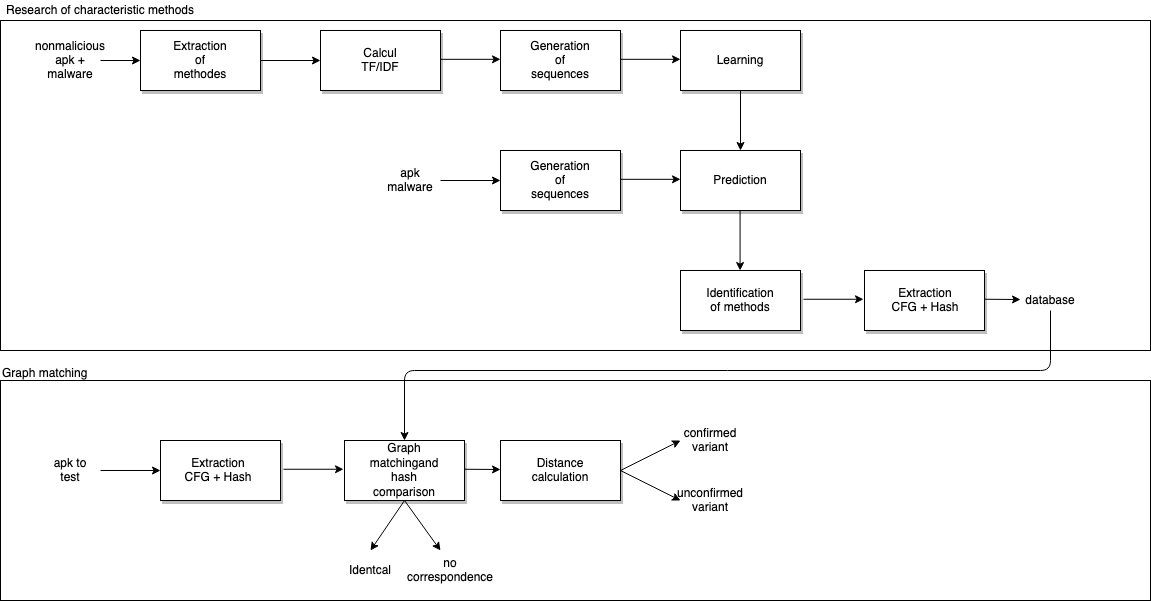} 
  \caption{General presentation of our approach}
  \label{global}
\end{figure}

Figure~\ref{global} presents the different steps of our approach. Due to a limited space, details of our approach are omitted here.

In order to detect if a software is infected, our method relies on the comparison between that software and a malware database. Control Flow Graphs (CFGs) are created based on the machine code of the software to be scanned. The structure of the software and its logical sequences is therefore preserved. The graph nodes are labelled with the hash of the underlying opcode block. 

The comparison between the CFGs of the software to be scanned and those of the database is carried out by calculating their subgraph isomorphisms. 
The calculation includes the structure of the graph and the correspondence of hashes. 
The computation for each graph pair being compared yields the following result: either the two CFGs are identical, which lets us know that the software under analysis is a piece of known malware; or the two CFGs do not match; or there is a match and in this case, the distance between the two CFGs is calculated to determine if it is a variant of a known malware family. 

The CFG database is critical to our method. This database is based on a database of known malware. The most characteristic CFGs get extracted for each type of malware. Evidently, the CFGs cannot match sequences of infected code, but they must allow for the best possible characterization of the malware in comparison with its other CFGs. The database is built using machine learning, which identifies the most characteristic machine code sequences of the malware. Once the characteristic machine code sequences have been identified, they will be transformed into CFGs to feed the CFG database.

As for any approach based on machine learning, the first step is learning a model from a relevant learning database, then detection can take place. Text mining is used in the learning process. The relative frequency of opcode ngrams is used to discriminate the blocks of infected code. Ngrams are short sequences of
opcodes.

Thus, learning is done through characteristic vectors from the analysis of the ngrams TF/IDF frequency in the software. 
Once the model has been learned from a database of clean software and malware for which the opcode TF/IDF frequency has been extracted, the sequences of infected code can be predicted in known malware. Each of those sequences is then transformed into a CFG in order to feed the database of CFGs characteristic of malware.

\section{Experiments}

Regarding the process of our method, as~\fig{global} presents, there is two main experimentation to perform: validating the CFG database machine learning method to choose; experimenting our approach on malware characterization. For this last experimentation step, we test our research work both on its theoretical background and on real data.

\subsection{Validating the CFG database machine learning's method}

As presented upper, the CFG database is built using machine learning, which identifies the most characteristic machine code sequences of the malware. We experiment the following supervised machine learning algorithms: Random Forest, XGBoost, Decision Tree, SVM and KNN, as they are well suited to class prediction from a database. The computations have been automated using the Dataiku tool. These algorithms were tested for ngrams of size 1 to 9 with the following F1 score as a result. The database uses 80\% learning database and 20\% tests. A 10-fold cross-validation is used for all the tests. 

\begin{table}[h]
 \hspace*{-0.5in}
\begin{tabular}{llllllllll}
                                          &                                     &                                     & \multicolumn{3}{l}{ngrams sizes}                                                                           &                                     &                                     &                                     &                                     \\ \cline{2-10} 
\multicolumn{1}{l|}{Algorithmes}          & \multicolumn{1}{l|}{1}              & \multicolumn{1}{l|}{2}              & \multicolumn{1}{l|}{3}              & \multicolumn{1}{l|}{4}              & \multicolumn{1}{l|}{5}              & \multicolumn{1}{l|}{6}              & \multicolumn{1}{l|}{7}              & \multicolumn{1}{l|}{8}              & \multicolumn{1}{l|}{9}              \\ \hline
\multicolumn{1}{|l|}{Random Forest}       & \multicolumn{1}{l|}{\textbf{0.923}} & \multicolumn{1}{l|}{\textbf{0.986}} & \multicolumn{1}{l|}{\textbf{0.935}} & \multicolumn{1}{l|}{\textbf{0.882}} & \multicolumn{1}{l|}{\textbf{0.846}} & \multicolumn{1}{l|}{\textbf{0.829}} & \multicolumn{1}{l|}{\textbf{0.820}} & \multicolumn{1}{l|}{\textbf{0.794}} & \multicolumn{1}{l|}{\textbf{0.796}} \\ \hline
\multicolumn{1}{|l|}{XGBoost}             & \multicolumn{1}{l|}{0.919}          & \multicolumn{1}{l|}{0.984}          & \multicolumn{1}{l|}{0.934}          & \multicolumn{1}{l|}{0.882}          & \multicolumn{1}{l|}{\textbf{0.846}} & \multicolumn{1}{l|}{\textbf{0.829}} & \multicolumn{1}{l|}{\textbf{0.820}} & \multicolumn{1}{l|}{0.793}          & \multicolumn{1}{l|}{\textbf{0.796}} \\ \hline
\multicolumn{1}{|l|}{Decision Tree}       & \multicolumn{1}{l|}{0.916}          & \multicolumn{1}{l|}{0.972}          & \multicolumn{1}{l|}{0.925}          & \multicolumn{1}{l|}{0.867}          & \multicolumn{1}{l|}{0.835}          & \multicolumn{1}{l|}{0.821}          & \multicolumn{1}{l|}{0.813}          & \multicolumn{1}{l|}{0.787}          & \multicolumn{1}{l|}{0.793}          \\ \hline
\multicolumn{1}{|l|}{KNN} & \multicolumn{1}{l|}{0.916}          & \multicolumn{1}{l|}{0.982}          & \multicolumn{1}{l|}{0.934}          & \multicolumn{1}{l|}{0.882}          & \multicolumn{1}{l|}{\textless 0.5}  & \multicolumn{1}{l|}{\textless 0.5}  & \multicolumn{1}{l|}{\textless 0.5}  & \multicolumn{1}{l|}{\textless 0.5}  & \multicolumn{1}{l|}{\textless 0.5}  \\ \hline
\multicolumn{1}{|l|}{SVM}                 & \multicolumn{1}{l|}{0.917}          & \multicolumn{1}{l|}{\textbf{0.986}} & \multicolumn{1}{l|}{\textbf{0.935}} & \multicolumn{1}{l|}{0.882}          & \multicolumn{1}{l|}{\textbf{0.846}} & \multicolumn{1}{l|}{\textbf{0.829}} & \multicolumn{1}{l|}{\textbf{0.820}} & \multicolumn{1}{l|}{0.793}          & \multicolumn{1}{l|}{\textbf{0.796}} \\ \hline
\end{tabular}
 \caption{F1 score for algorithms based on ngram size for vector size of 100} \label{tabl}
\end{table}

As Table~\ref{tabl} presents, Random Forest and SVM show the best F1 scores. Conversely, KNN has rapidly lost in efficiency as soon as the ngram size reached 5 or above. Regarding the F1 scores of the best algorithms, we can assume that the present approach is able to identify features of interest among the opcodes of several hundred of apks so that a database can be built to search for isomorphisms.

\subsection{Laboratory database}

In this section, the objective is to test the performance and the robustness of the method. A common way to create malwares is to add infected codes to an existing Android software. However, existing works doesn't focus on verifying if only the malware is detected as one and not the safe original application. For this purpose we create this laboratory database. Moreover, our ame is also to detect variants of the same malware. In order to do this database, we introduced three variants of an infected code in ten Android applications, thereby creating the equivalent of 30 infected applications, all being variants of the same malware. The database was completed with the 10 original and unmodified applications and 100 other applications. The laboratory testing allowed us to validate the underlying concept of our approach.

The main benefit of this laboratory database is that one can dissect the results variant by variant. Three dictionaries were thus created, each one based on a unique variant. Table~\ref{tab-labo} shows the results obtained using our method for each of the three dictionaries: all variants are found and their is no false-positive. The match threshold used to get those results was that a subgraph isomorphism was indeed found in the candidate program and at least half the hashes were a match. The results fully validate the present method, though the choice of match threshold should not be too high.

\begin{table}
  \centering
  \begin{tabular}{|l|ccc|}
    \hline
    & Precision & Recall & F-mesure \\
    \hline
    Variant 1 & 1 & 1 & 1 \\
    Variant 2 & 1 & 1 & 1 \\
    Variant 3 & 1 & 1 & 1 \\
    \hline
  \end{tabular}
  \caption{Results obtained on the laboratory database}
  \label{tab-labo}
\end{table}

\subsection{Set of real data}

In this section, we test the present method on a set of real data. In total, 10 variants of Droidjack and Opfake malware were collected and a database of 100 clean programs and as much malware was built. Thus, our method has been tested in real conditions.

In order to validate the method in real conditions, a database of 100 completely clean applications and 100 malware has been created. The infected specimens were collected from the collaborative platform Koodous~\cite{koodous}. Among those specimens, two in particular, DroidJack and Opfake, were selected as well as their variants - 6 DroidJack variants and 4 Opfake variants were collected. 

Three tests were carried out with a dictionary created: based solely on a DroidJack variant; based solely on an Opfake variant; based on all the 100 malware.

Table~\ref{tab-reel} introduces the results of those three tests. The test carried out on the 6 known DroidJack variants shows that there were two false positives. However, the CFGs of those two false positives and those of Droidjack are a perfect match, which could mean that the two false positives may have been misfiled in Koodous.

\begin{table}
  \centering
  \begin{tabular}{|l|ccc|}
    \hline
    Dictionary based on & Precision & Recall & F-mesure \\
    \hline
    DroidJack & 0,98 & 1 & 0,99 \\
    Opfake & 1 & 1 & 1 \\ 
    Set of malware & 1 & 1 & 1 \\
    \hline
  \end{tabular}
  \caption{Results obtained on the real data set consisting of 100 healthy applications and 100 malware}
  \label{tab-reel}
\end{table}

\section{Conclusion}

This paper presents a malware detection method based on the construction of a control flow graph from Android bytecode mnemonics. The characterization of new malware is carried out by comparing its signature with that of known malware using subgraph isomorphism. Reference signatures are identified by a machine learning process. The results obtained from an ad hoc laboratory database and a set of real data validate the present approach with an almost perfect detection rate. Due to a limited space, details of our approach will be presented in an extended work. More experiments on other and more large database will be performed.

\bibliographystyle{plainurl}
\bibliography{Android_malware_detection}

\end{document}